\documentclass[12pt,letterpaper]{article}

\usepackage{amsmath,amssymb,calc}
\usepackage{graphicx}

\newcommand\be{\begin{equation}}
\newcommand\ee{\end{equation}}
\newcommand{\bea}{\begin{eqnarray}}
\newcommand{\eea}{\end{eqnarray}}

\newcommand{\hlf}{\frac{1}{2}}

\newcommand{\C}[1]{$(\ref{#1})$}

\newcommand\zero[1]{^{(0){#1}}}

\def\tr{{\rm tr\,}}

\newcommand{\ap}{\alpha'}

\newcommand{\nn}{\nonumber}
\newcommand{\pd}{\partial}

\def\id{\protect{{1 \kern-.28em {\rm l}}}}

\def\cN{{\cal N}}
\def\N{{\cal N}}
\def\cV{{\cal V}}

\def\cM{{\cal M}}

\def\D{\Delta}

\def\la{\lambda}

\def\dd{\delta}
\def\del{\partial}
\def\a{\alpha}
\def\b{\beta}
\def\g{\gamma}

\def\g{\gamma}

\def\k{\kappa}

\def\o{\omega}

\newcommand{\jbar}{{\bar \jmath}}
\def\1{^{(1)}}
\def\0{^{(0)}}
\def\2{^{(2)}}

\def\ie{i.e.}

\font\mybb=msbm10 at 12pt

\def\bb#1{\hbox{\mybb#1}}

\def\Z {\bb{Z}}

\newcommand{\R}{{\mathbb R}}

\def\K{K\"ahler}

\def\id{\protect{{1 \kern-.28em {\rm l}}}}

\let\non\nonumber

\begin{document}
\begin{titlepage}

\begin{center}

January 25, 2011 \hfill         \phantom{xxx} \hfill EFI-10-16

\vskip 2 cm {\Large \bf Quantum Corrections to Heterotic Moduli Potentials}

\vskip 1.25 cm {\bf  Lilia Anguelova$^{a}$\footnote{anguella@ucmail.uc.edu} and Callum Quigley$^{b}$\footnote{cquigley@uchicago.edu}}\non\\
{\vskip 0.5cm  $^{a}$ {\it Dept. of Physics, University of Cincinnati,
Cincinnati, OH 45221, USA}\non\\ \vskip 0.2 cm
$^{b}$ {\it Enrico Fermi Institute, University of Chicago, Chicago, IL 60637, USA}\non\\}

\end{center}
\vskip 2 cm

\begin{abstract}
\baselineskip=18pt

In a recent paper, we derived the leading $\alpha'$ corrections to the K\"ahler potentials for moduli in $(0,2)$ heterotic compactifications. In the same spirit as the LARGE volume stabilization scenario for type IIB orientifolds, we examine whether these quantum corrections, together with a combination of tree-level and non-perturbative superpotentials, are sufficient to stabilize the overall volume modulus at large values. This is not a priori obvious, since the corrections we found are of a lower order than those used in the type IIB setting. Nevertheless, we find that stabilizing the volume at (exponentially) large values may be possible (under certain conditions) in these heterotic backgrounds.

\end{abstract}

\end{titlepage}


\section{Introduction}

Stabilizing the moduli of string compactifications is of prime interest for phenomenology, since their values determine the various parameters of the four-dimensional
effective description. In recent years there has been a lot of progress in this direction, based on the realization of \cite{DRS,GVW} that non-vanishing
background fluxes lead to nontrivial moduli potentials.\footnote{For a comprehensive review of flux compactifications, see \cite{MG}.} In some classes of
string compactifications it is even possible to stabilize all geometric moduli via the flux-induced superpotential.\footnote{This is the case when the superpotential
depends on all of those moduli, as in type IIA with all RR fluxes turned on \cite{IIA}, M-theory on $SU(3)$ structure manifolds \cite{Mth} or type IIA/B on
$SU(3)\times SU(3)$ structure \cite{BG}.} Such a stabilization at the classical level would seem rather desirable conceptually. However, so far the most
phenomenologically promising compactifications are those in which quantum effects are used to stabilize at least some of the moduli. For example, this is
the case for the type IIB large volume compactifications of \cite{BBCQ} and the M-theory $G_2$-MSSM model of \cite{Kane}.

The above situation is acutely exemplified by the heterotic string. Namely, compactifications on non-K\"{a}hler manifolds lead to a flux superpotential for
all geometric moduli \cite{BBDP,HetW}. However, the resulting stabilization is such that there are typically string-scale cycles \cite{BBDP,BBDG}, which
raises questions about the reliability of the supergravity approximation within which the 4d effective potential was derived. Interestingly enough, until
now the possibility of stabilizing some of the CY moduli via $\alpha'$-corrections (together with other quantum effects\footnote{The role of world-sheet
instantons for moduli stabilization in the heterotic string has been investigated in \cite{CKL}; see also \cite{BKOS} and references therein for a study of worldsheet instantons in the strongly coupled limit, i.e. in heterotic M-theory.}) has not been pursued in the context of heterotic compactifications. We will study precisely this
issue. In particular, we will investigate quantum effects that induce a potential for the K\"{a}hler moduli, while viewing the complex structure ones as stabilized
by the flux superpotential, just like in the type IIB compactifications of \cite{KKLT,BBCQ}.

The leading $\ap$-corrections to the moduli \K\ potentials in $(0,2)$ compactifications were worked out recently by the authors and S. Sethi in \cite{AQS}. We
review these results in Section 2. In Section 3, we begin investigating
the role they play in stabilizing moduli. Specifically, we include world-sheet
instanton corrections to the classical flux superpotential and then study the interplay, within the resulting scalar potential, between these non-perturbative
effects in $W$ and the perturbative $\ap$-corrections to $K$. Our goal will be to find out whether it is possible to have large volume minima in the vein of \cite{BBCQ}. By this we mean that the volume is stabilized in such a way, that its value at the minimum is proportional to the exponential of a two-cycle volume. For more clarity, we consider first the case of only two K\"{a}hler moduli. However, it turns out that in this case there are no stable minima of the desired type. Then, in Section 4, we show that for at least three K\"{a}hler moduli, and under a certain additional condition, it is possible to find non-supersymmetric minima at large volume (in string units)
in the vein of \cite{BBCQ}. Such large volume minima should be very promising phenomenologically and could have far reaching consequences for the kinds of predictions
that one can extract from heterotic compactifications. Finally, in Appendix \ref{SpHerm} we collect some more technical details about Special Hermitian geometries and in Appendix \ref{NoScaleBreaking} we give the detailed computation of the leading $\alpha'$ correction to the scalar potential.

\section{The $(0,2)$ K\"{a}hler Potential}\label{(2,2)review}
\setcounter{equation}{0}

In reducing the heterotic string to four dimensions on a Calabi-Yau background $\cM$, a number of massless moduli fields appear in the low-energy $\N=1$
supergravity effective action. In addition to the axio-dilaton there are the \K, complex structure, and gauge bundle moduli. For the applications in the next section,
we will be primarily interested in the K\"{a}hler moduli, which we will denote by
\be
T^\a = b^\a + i t^\a.
\ee
We will ignore the rest of the moduli fields, assuming that they are already fixed by the flux superpotential.
The imaginary parts of $T^\a$ parameterize the \K\ form of the internal space $J = t^{\a} \o_\a$, where $\o_\a$ form a basis for
$H^2(\cM,\Z)$, and the $b^\a$ are the associated axions that arise from the NS B-field. Classically, the action for these fields is controlled by the \K\ potential\footnote{For simplicity,
we set the four dimensional Newton's constant $\k^2$ to 1.}
\be
K = -\log\cV
\ee
where the volume $\cV$ of the internal space is given by
\be
\cV = {1\over6}\int_\cM J\wedge J\wedge J = {1\over6}\k_{\a\beta\gamma}t^\a t^\beta t^\gamma,
\ee
and $\k_{\a\beta\gamma} = \int_\cM \o_\a\wedge \o_\beta\wedge \o_\gamma$ are the triple intersection numbers of $\cM$.
When the gauge bundle is chosen to satisfy the standard embedding (\ie\ when the worldsheet theory is a (2,2) SCFT), the leading $\ap$ correction appears
at $O(\ap^3)$ and the \K\ potential is corrected simply by replacing the volume by the ``quantum corrected volume" $\cV\mapsto \cV + \ap^3\zeta(3)\chi$,
where $\chi$ is the Euler character of $\cM$ \cite{COGP}.

However, our interests lie in the more general class of theories, where the bundle does not satisfy the standard embedding, and so the worldsheet theory
has only $(0,2)$ supersymmetry. As shown originally in \cite{EW,WW} (and again more recently in \cite{GPT}) the zeroth order Calabi-Yau metric is no longer
a solution to the $\ap$-corrected equations of motion. Instead, one finds a new non-K\"{a}hler metric
\be
g'_{i\jbar} = g_{i\jbar} + \ap h_{i\jbar} +\ldots
\ee
where $g_{i\jbar}$ is the original Calabi-Yau metric, and the non-\K ity\footnote{The fact that $h_{i\jbar}$ in non-\K\ is most easily understood by considering
its associated 2-form, which we will simply denote by $h$. Then in \C{hsol} one can replace $\D_L$ by the usual Laplace-Beltrami operator $\D$, so that
$\D h \sim F^2 -R^2\neq0$. Since $h$ is not harmonic, it is not closed. Thus $h$ is not \K.} stems from the correction term
\be\label{hsol}
h_{i\jbar} = \hlf\D_L^{-1} \Big[\tr\big(F_{i\bar{k}}F_\jbar{}^{\bar{k}}\big) - R_{i\bar{k}\ell\bar{m}}R_\jbar{}^{\bar{k}\ell\bar{m}}\Big].
\ee
Here $\D_L$ is the Lichnerowicz operator (a certain 2nd order differential operator, whose precise form can be found in the references) associated to the metric
$g$.\footnote{The zeroth order metric $g$ is also used to raise and lower indices.} We derived the effect of this metric
deformation on the classical \K\ potential in \cite{AQS}, to which we refer the reader for more details. The
net result is quite similar to the $(2,2)$ case, namely the classical volume gets replaced by a quantum corrected volume:
\be
\cV \mapsto \cV - {\ap^2\over2} \big(h,h\big) + O(\ap^3).
\ee
The inner-product $(\,\,,\,)$ has several equivalent presentations, depending on whether one views $h_{i\jbar}$ as the components of a 2-tensor,
or of a $(1,1)$-form:
\be
\big(h,h\big) = \int_{\cM} \sqrt{g}\,\,  h_{i\jbar}\,h^{i\jbar} = \int_{\cM} h\wedge* h = -\int_{\cM} J\wedge h \wedge h.
\ee
In writing the last equality\footnote{Recall that for a two form $\o$, its Hodge dual is $*\o = -J\wedge\o + {3\over2}{\int J\wedge J\wedge \o \over \int J \wedge J \wedge J} J\wedge J$.},
we made use of the fact that for \C{hsol} to be well-defined $h_{i\jbar}$ must be orthogonal to the zero-modes of Lichnerowicz
(so that we can invert $\D_L$), which in particular means that
\be
0 = (g,h) = \int_{\cM}\sqrt{g}\, g^{i\jbar}h_{i\jbar} = \hlf\int_{\cM} J\wedge J\wedge h .
\ee

Finally, let us make a useful observation about the scaling behaviour of $g$ and $h$.
It is clear that the Calabi-Yau metric $g$ scales
with the volume as $\cV^{1/3}$, and since the perturbative expansion is not just in $\ap$, but rather in $\frac{\alpha'}{{\cal V}^{1/3}}$, then $h$ should not
scale with the overall volume of $\cM$. More precisely, we can always write:
\be
g'_{i\jbar} = \cV^{1/3}\tilde{g}_{i\jbar} + \ap h_{i\jbar} + \ldots  ,
\ee
where $\tilde{g}$ and $h$ are invariant under scaling of $\cV$. Furthermore, one can see from (\ref{hsol}) that $h$ is actually invariant under a {\it uniform} scaling of the K\"{a}hler moduli $t^\a\rightarrow\la t^\a$, \ie\ it is a homogeneous function in $t^\a$ of degree zero. This fact will be important later.


\section{The Scalar Potential and Moduli Stabilization} \label{ModStab}
\setcounter{equation}{0}

In this section we begin studying the implications of the $\alpha'$-corrected K\"{a}hler potential for moduli stabilization. We will consider the scalar potential
of the effective 4d $\cN=1$ theory. As is well-known, it is completely determined by the superpotential $W$ and the K\"{a}hler potential $K$. Let us start by specifying
$W$. In heterotic compactifications, the complex structure moduli enter the superpotential via the GVW expression \cite{BC}:
\be \label{Wfl}
W_{flux} = \int_{\cM} H \wedge \Omega .
\ee
To stabilize the K\"{a}hler moduli, one needs to either take into account non-perturbative effects or consider non-K\"{a}hler compactifications. In the latter
case, $W$ becomes ~\cite{BBDP,HetW} $\int (H + idJ)\wedge \Omega$, where clearly the K\"{a}hler moduli enter through $dJ$. However, as we pointed
out in the Introduction, in non-K\"{a}hler compactifications one usually ends up with cycles whose volumes are of order $\alpha'$. Hence the supergravity
approximation, within which these minima were derived, is not reliable. So we will focus on non-perturbative effects in the following. Another, perhaps the
most important, motivation for this is to investigate whether one can find minima at exponentially large volume\footnote{We mean large in string units.},
similar to the type IIB large volume scenarios of \cite{BBCQ}.\footnote{The IIA version of large volume compactifications was constructed in \cite{PTW}}
The key feature of these compactifications is the balancing between nonperturbative effects in $W$, due to brane instantons or gaugino condensation, and
$\alpha'$ corrections in the K\"{a}hler potential. We want to investigate whether the same kind of balancing can occur in the present case, given the $\ap$
corrections of the previous section.

Let us pause for a moment to discuss the consistency of this approach. After all, we have seen that a $(0,2)$  Calabi-Yau background is deformed into a
non-\K\ manifold at $O(\ap)$. One could be worried that the superpotential contribution $\int dJ \wedge \Omega$, coming from the
``geometric flux" would dominate over any non-perturbative terms. However, it turns out that
the geometries involved in our compactifications are of a very special type, known as \textit{Special Hermitian}: these are non-K\"{a}hler manifolds (so $dJ \neq 0$), but
$dJ \wedge \Omega = 0$ identically. The backgrounds we consider here are precisely within this class of compactifications,
since the dilaton is constant to ${O} (\alpha'^2)$ \cite{AQS}.\footnote{For more details on these geometries, see Appendix \ref{SpHerm}.}
Also, it is clear from the non-renormalization theorems for the superpotential that
perturbative $\ap$ corrections cannot alter $W$, so it had better be the case that $\int dJ \wedge \Omega$ vanishes in our backgrounds, since it does so at zeroth order. Thus,
non-perturbative effects are the only (known) method to stabilize \K\ moduli in this set-up. Let us now enumerate the possibilities.

\subsection{Non-perturbative superpotential}

The non-perturbative effects at our disposal are world-sheet instantons, gaugino condensation and NS5-brane instantons. Let us address them in turn. It has
been known since \cite{DSWW}
that in pure CY compactifications with the standard embedding (i.e., compactifications on $(2,2)$ conformal theories) the superpotential generated by
worldsheet instantons vanishes. In fact, no superpotential is generated by worldsheet instantons for a large class of $(0,2)$ compactifications,
namely those for which a GLSM description exists~\cite{NoW}. However, for a generic
$(0,2)$ background we do expect to have $W_{inst} \neq 0$. So we will consider a non-perturbative superpotential of the form:
\be \label{Winst}
W_{inst} = \sum_{\alpha} A_{\alpha} e^{i a_{\alpha} T^{\alpha}}  ,
\ee
where $T^{\alpha}$ are the complexified Kahler moduli of the 2-cycles wrapped by the world-sheet instantons and $A_{\alpha}$,
$a_{\alpha}$ are constants. One should note that, in principle, the coefficients $A_{\alpha}$ can depend on complex structure and gauge bundle moduli.
 However, for the present treatment we will view them as constants similar to \cite{KKLT,BBCQ}.

Another non-perturbative effect that introduces K\"{a}hler moduli dependence into the superpotential is gaugino condensation. Unlike the type II case however, for the heterotic string the resulting superpotential has a functionally different form compared to the worldsheet-instanton generated expression (\ref{Winst}). Namely, the gaugino condensation superpotential is:
\be \label{WGC}
W_{GC} = A e^{ia(S+\beta_{\alpha} T^{\alpha})} ,
\ee
where the $\beta_{\alpha} T^{\alpha}$ term arises from one-loop threshold corrections and $a$, $A$ are constants. For a recent study of moduli stabilization with (\ref{WGC}), see \cite{GKLM}.
The moduli stabilization mechanisms for $W_{inst}\neq 0$ and for $W_{GC} \neq 0$ are rather different. In particular, gaugino condensation cannot lead to a large volume minimum in the vein of \cite{BBCQ}.
The underlying reason is that, while in (\ref{Winst}) the dominant term is the one with the smallest cycle just like for $W_{np}$ studied in \cite{BBCQ},
the form of (\ref{WGC}) is such that the dominant contribution comes from the cycle with the largest volume. The relation between this dominant behaviour
and the presence or absence of a large volume minimum will become more clear below. So we will use $W_{inst}$, instead of $W_{GC}$, as the source of
K\"{a}hler moduli dependence in the superpotential. One could still keep the leading contribution from gaugino condensation\footnote{Generically, the one-loop
coefficients $\beta_{\alpha}<\!\!<1$, although it is possible to find examples in which $\beta_{\alpha}\approx {O}(1)$, as pointed out in \cite{GKLM}.}
\be
W_{GC}' = A e^{iaS},
\ee
in order to stabilize the dilaton. In such a case, we will have a superpotential
\be
W' = W_{flux} + W_{GC}'
\ee
which is responsible for freezing the complex structure moduli as well as the axio-dilaton. Note that in order to fix $S$ at a finite supersymmetric value,
$D_S W=0$ requires $W_{flux}\neq0$, despite the fact that this would break supersymmetry in the absence of a gaugino condensate. These observations are consistent
with \cite{GKLM}.
In the following, we will assume that such a supersymmetric stabilization of $S$ and the complex structure has already been performed
and concentrate on the potential for the remaining moduli, namely the K\"{a}hler ones. As already mentioned, our goal will be to search for large volume minima arising from the interplay of $\alpha'$ corrections and world-sheet instantons.

Before turning to that, let us briefly mention the third kind of non-perturbative effects available in the present context, namely NS5-brane instantons. The latter have been considered for moduli stabilization purposes in type II compactifications in \cite{LV}.
However, since the superpotential they induce is of the form
\be
W_{NS5} = const \times e^{- const' \times {\cal V}}  ,
\ee
where ${\cal V}$ is the overall CY volume modulus, they also do not lead to large volume minima, just like gaugino condensation does not. In fact, $W_{NS5}$ is obviously completely negligible at large volume.

To recapitulate, we will consider the standard 4d $\cN=1$ scalar potential
\be \label{ScPot}
V = e^K \left( G^{\alpha \bar{\beta}} D_{\alpha} W \bar{D}_{\bar{\beta}} \bar{W} - 3 |W|^2 \right)  ,
\ee
where $\alpha$, $\bar{\beta}$ run over the K\"{a}hler moduli,
\be \label{Wtot}
W = W_0 + W_{inst}
\ee
with $W_0 = \langle W' \rangle$ being a constant that results from the frozen dilaton and complex structure
moduli, and finally
\be \label{Kkm}
K = -\ln {\cal V} + {\alpha'^2\over2}\, \frac{(h,h)}{\cV} +O(\ap^3) ,
\ee
where we have kept only the leading ${O} (\alpha'^2)$ correction to the K\"{a}hler potential. We have dropped additive constants from $K$
that originate in the frozen moduli sectors, as these only lead to an overall rescaling of $V$.

\subsection{Scalar potential at large volume} \label{LVPot}

We will look for a large-volume nonsupersymmetric minimum of the scalar potential (\ref{ScPot}). As in
\cite{BBCQ},
we can write this potential in the following manner:
\be \label{Vtot}
V = V_{np\,1} + V_{np\,2} + V_{\alpha'}  ,
\ee
where $V_{\alpha'}$ is the leading $\alpha'$ correction that we will turn to a bit later and
\be \label{Vnp1i}
V_{np\,1} = e^K G^{\a \bar{\beta}} \left( a_{\alpha} A_{\alpha} a_{\beta} \bar{A}_{\beta}  e^{i(a_{\alpha} T^{\alpha} - a_{\beta} \bar{T}^{\beta})} \right)  ,
\ee
while
\be \label{Vnp2i}
V_{np\,2} = i e^K \left( G^{\alpha \bar{\beta}} a_{\alpha} A_{\alpha} e^{i a_{\alpha} T^{\alpha}} \bar{W}_0 \pd_{\bar{\beta}} K - G^{\beta \bar{\alpha}} a_{\beta} \bar{A}_{\beta} e^{-i a_{\beta} \bar{T}^{\beta}} W_0 \pd_{\alpha} K \right)  .
\ee
Since our non-perturbative superpotential has the same form as the one in
\cite{BBCQ},
the above expressions are also exactly the same as those in the IIB orientifold case. The difference is in the meaning of the variables $T^{\alpha}$:
here they correspond to the (complexified) two-cycle volumes, whereas in \cite{BBCQ}
they correspond to four-cycles. Let us now establish some notation for the CY volume and its derivatives, in terms of the (real) 2-cycle volumes $t^{\alpha}$:
\bea \label{RelVt}
{\cal V} =& \frac{1}{6} \kappa_{\alpha \beta \gamma} t^{\alpha} t^{\beta} t^{\gamma}& \non\\
{\cal V}_{\a} =& \frac{1}{2} \kappa_{\alpha \beta \gamma} t^{\beta} t^{\gamma} &\\
{\cal V}_{\a\beta} =& \kappa_{\alpha \beta \gamma}  t^{\gamma} &\non\\
{\cal V}^{\a\beta} =& \big(\cV_{\a\beta}\big)^{-1}&.\non
\eea
To compare our results with the IIB literature we note that the 4-cycle volumes are given by $\tau_{\alpha} =  {\cal V}_\a$,\footnote{To be precise,
in mapping to IIB variables we should multiply the $t^{\a}$ by $e^{\Phi/2}$, with $\Phi$ being the 10d dilaton field \cite{GKP,BBHL}. Since the dilaton
is typically frozen in IIB flux compactifications, we may neglect this rescaling.} and
the corresponding complex moduli are $\rho_{\alpha} = \sigma_{\alpha} + i \tau_{\alpha}$ with $\sigma_{\alpha}$ being axions.

Now, we will extract the large-${\cal V}$ behaviour of $V_{np\,1}$ and $V_{np\,2}$ in the same manner as in \cite{BBCQ}. To do that, one has to make two basic assumptions. The first is that there is a set of two-cycles, whose volumes $t_s^1, ... , t_s^m$ remain much smaller than the volumes of all the other 2-cycles in the limit ${\cal V} \rightarrow \infty$. And the second is that in this limit the CY volume ${\cal V}$ is of the same order of magnitude as a power of $e^{a_k t^k_s}$ for every $k = 1,..., m$. The role of the second assumption will become more clear below. For more clarity, we will consider here the case of a single small modulus, that we denote just by $t_s$. However, we will find out that for the existence of large volume minima we need at least two small moduli; the latter case will be studied in the next section.

As a result of the first assumption above, in the large volume limit exponentials of all other $t^{\alpha}$ can be neglected in comparison to those of $t_s$. Hence, in that limit we find:
\be \label{Vn1}
V_{np\,1} = e^K G^{s \bar{s}} a_s^2 |A_s|^2 e ^{-2 a_s t_s} \nn \\
\ee
and
\be \label{Vn2}
V_{np\,2} = 2 \,e^K G^{s \bar{\beta}} a_s |A_s| |W_0| e^{-a_s t_s} K_{\beta}  ,
\ee
where we have also minimized with respect to the axions.\footnote{This minimization is, clearly, exactly the same as in the IIB large volume literature; for detailed treatment see appendix A of
\cite{CCQ}.
The upshot is that $a_s b_s = n \pi - \theta_s + \theta_W$, where $n \in 2 \mathbb{Z} + 1$ and $\theta_s$, $\theta_W$ are the phases of $A_s$ and $W_0$ respectively.}
To compute the inverse metric $G^{\a\bar{\beta}}$, note that in the large ${\cal V}$ limit the moduli space metric can be written as:
\be \label{G-leading}
G_{\a \bar{\beta}} = -{\cV_{\a\beta}\over\cV} + {\cV_\a \cV_\beta \over\cV^2} + O(\ap^2) .
\ee
Hence, one can easily see that
\be \label{G-inv}
G^{\a\bar{\beta}} = -\cV \cV^{\a\beta} + \hlf t^\a t^\beta\, + O(\ap^2),
\ee
where we have used the fact that $\cV^{\a\beta}\cV_\a = \hlf t^\beta$.  Using this expression, we can determine the non-perturbative contributions to the
scalar potential, beginning with $V_{np\,2}$. Defining $\mu = 2 a_s |A_s| |W_0|$, we find from \C{Vn2}:
\bea \label{Vnp2}
V_{np\,2} &=& \mu e^K G^{s\beta}e^{-a_s t_s} K_\beta \non\\
&=& {-\mu\over \cV^2}\left(-\cV \cV^{s\beta}+\hlf t_s t^{\beta}\right) e^{-a_s t_s}\cV_\beta + \ldots\\
&=& -\mu {t_s e^{-a_s t_s}\over\cV} + O\left({e^{-a_s t_s} \over \cV^2}\right) , \non
\eea
where $...$ denotes the subleading terms, and we used the relation $t^\a \cV_\a = 3\cV$. Now, turning to $V_{np\,1}$, from \C{Vn1} we have:
\bea
V_{np\,1} &=& e^K G^{ss} a_s^2 |A_s|^2 e^{-2a_s t_s} \non\\
&=&\left(-\cV^{ss} + {t_s^2\over 2\cV}\right) a_s^2|A_s|^2 e^{-2a_s t_s} +\ldots .
\eea
To understand the large ${\cal V}$ behaviour of $\cV^{ss}$, let us now for simplicity concentrate on the case of only two K\"{a}hler moduli;
we will denote them by $t_s$ and $t_\ell$, where $t_\ell\gg t_s$ in the large volume limit. As we will recall below, this case already captures all the
essential features of the IIB large volume scenario, whose analogue in the present context we are after. With only two moduli,
\be
\cV^{ss} = \frac{\kappa_{\ell\ell s} t_s + \kappa_{\ell\ell\ell} t_\ell}{(\kappa_{sss} t_s + \k_{ss\ell} t_\ell) (\kappa_{\ell\ell s} t_s + \kappa_{\ell\ell\ell} t_\ell) - (\kappa_{s\ell s} t_s +\kappa_{s\ell\ell} t_\ell)^2} .
\ee
We will consider only the generic case here, i.e. with all components of $\k_{\a\beta\g}$ non-vanishing.
In the limit $t_\ell\gg t_s$, the volume is\footnote{Clearly, other behaviours are possible in special cases, such as $\kappa_{\ell \ell \ell} = 0$ while $\kappa_{s \ell \ell} \neq 0$ resulting in $\cV \sim t_{\ell}^2$. We leave those for future investigation.}
\be
\cV \sim {1\over6}\k_{\ell\ell\ell}\,t_\ell^3+\ldots\,
\ee
and generically
\be
\cV^{ss} \sim {\k_{\ell\ell\ell}\over \k_{ss\ell}\,\k_{\ell\ell\ell} - \k_{s\ell\ell}^2} t_\ell^{-1} \sim{\k_{\ell\ell\ell}^{4/3}\over \k_{ss\ell}\,\k_{\ell\ell\ell} - \k_{s\ell\ell}^2} \big(6\cV\big)^{-1/3},
\ee
which leads to
\be \label{ldef}
V_{np\,1} = \la {e^{-2a_s t_s}\over \cV^{1/3}} + O\left({e^{-2a_s t_s}\over \cV}\right)
\ee
with $\la = a_s^2 |A_s|^2 \k_{\ell\ell\ell}^{4/3} / 6^{1\over3}(\k_{s\ell\ell}^2 - \k_{ss\ell}\,\k_{\ell\ell\ell} )$.

Now let us turn to the $\alpha'$ correction to the scalar potential $V_{\alpha'}$. It arises from the term $G^{\alpha \bar{\beta}} K_{\alpha} K_{\bar{\beta}} |W|^2$,
as this is precisely the term that cancels $-3|W|^2$ in the absence of quantum corrections and leads to the no-scale structure of the classical theory.
To extract the leading behaviour of $V_{\alpha'}$, let us first write the norm of $h$ as
\be\label{hh}
(h,h) = -\int_{\cM} J(t) \wedge h(t) \wedge h(t) =\, t^{\a} f_{\a}(t)\, \simeq\, t^\ell\, f_\ell(t) +\ldots
\ee
where
\be
f_{\a}(t) = -\int_{\cM} \o_{\a} \wedge h(t) \wedge h(t),
\ee
is a homogeneous degree zero function, because $h$ is. In the last step of \C{hh} we used the fact that we are considering the large volume limit $t_\ell \sim \cV^{1/3} \gg t_s$. Note that this definition implies that $f_\ell(t)\geq0$. In what follows, we will simply omit the $\ell$ subscript of $f_\ell\equiv f$, in which case the K\"{a}hler potential (\ref{Kkm}) can be written as
\be \label{Kexpr}
K = - \ln {\cal V} +  {\alpha'^2\over2} \frac{f(t)}{{\cal V}^{2/3}} .
\ee
Now, it is easy to compute the leading $\alpha'$ contribution to the scalar potential (we leave the details to Appendix \ref{NoScaleBreaking}):
\be \label{Valp}
V_{\alpha'} = e^K \left( G^{\alpha \bar{\beta}} K_{\alpha} K_{\bar{\beta}} |W|^2 - 3 |W|^2 \right) = -\nu \,\frac{f(t)}{{\cal V}^{5/3}} + \,{O} \!\left( \frac{1}{{\cal V}^{8/3}} \right) ,
\ee
where $\nu = {\ap^2} |W_0|^2 ({6\over \k_{\ell\ell\ell}})^{1/3}$ is a constant. This result should be contrasted with
the analogous situation in type IIB orientifolds, where $V_{\ap}$ vanishes at this order in $\ap$ \cite{AQS,CCQ2}.

\subsection{Quest for large volume minima}

So far, we extracted the leading large volume behaviour of each of the terms in (\ref{Vtot}). For comparison with the IIB orientifold case, let us recall
that at large volume the IIB scalar potential has the following form \cite{BBCQ}:
\be \label{LVCIIB}
V = \hat{\lambda} \frac{\sqrt{\tau_s} e^{-2 a_s \tau_s}}{{\cal V}} - \hat{\mu} \frac{\tau_s e^{-a_s \tau_s}}{{\cal V}^2} + \frac{\hat{\nu}}{{\cal V}^3}  ,
\ee
where $\hat{\lambda}$, $\hat{\mu}$ and $\hat{\nu}$ are (positive) constants and, for simplicity, we are again considering only two K\"{a}hler moduli; this potential
captures all the essential features of the IIB large volume scenario. Obviously, for ${\cal V}\sim e^{a_s \tau_s}$ all three terms above are of the same
order of magnitude. Furthermore, all the subleading terms, that were neglected in each of $V_{np1}$, $V_{np2}$ and $V_{\alpha'}$, are higher order compared
to (\ref{LVCIIB}) and so the expansion is consistent. This is the reason for the second assumption we mentioned in the beginning of Section \ref{LVPot},
namely that ${\cal V}$ is of the same order of magnitude as a power of the exponential of the small cycle. Now, extremizing (\ref{LVCIIB}) with respect to
${\cal V}$ and $\tau_s$, one finds the large volume minimum of \cite{BBCQ}.

To reproduce the above situation, we need to see whether in our case it is possible to have comparable contributions from $V_{np1}$, $V_{np2}$ and $V_{\alpha'}$ for
volume modulus of the order ${\cal V}\sim e^{q a_s t_s}$ for some number $q$. It is easy to realize that this can be achieved for
${\cal V} \sim e^{\frac{3}{2} a_s t_s}$, when the function $f(t)$ does not introduce additional powers of $\cV$. This happens only for $f(t) = const$. Indeed, since $f$ is homogeneous of degree zero and we are considering the case of only two K\"{a}hler moduli here, then clearly $f(t) = f\left( \frac{t_s}{t_\ell} \right)$. Furthermore, in the large volume limit $t_{\ell}$ is given by $\cV^{1/3}$. Hence $f(t)$ is of the form $f\left( \frac{t_s}{\cV^{1/3}} \right)$.
So let us first consider $f(t) = const$.\footnote{We will turn to the case of nontrivial $f(t)$ shortly.} In such a case, the leading contribution in each term in \C{Vtot} is ${O} ({\cal V}^{-5/3})$ and so, at large volume, we have the following scalar potential:
\be\label{LVSPot}
V = \lambda \frac{e^{-2 a_s t_s}}{{\cal V}^{1/3}} - \mu \,\frac{t_s \,e^{- a_s t_s}}{{\cal V}} - \nu \,\frac{f(t)}{{\cal V}^{5/3}}  .
\ee
Note the qualitative similarity of the above expression to the IIB scalar potential (\ref{LVCIIB}).

Now let us see whether the potential (\ref{LVSPot}) can stabilize both $\cV$ and $t_s$. Since $f$ is constant here, we can be absorbed it in the definition of $\nu$. Then it is easy to see that the condition $\frac{\pd V}{\pd {\cal V}} = 0$ gives:
\be \label{Vsol}
{\cal V}^{2/3} = \frac{3}{2} \frac{\mu}{\lambda} t_s e^{a_s t_s} \left( 1 \pm \sqrt{1 + \frac{20}{9} \frac{\lambda \nu}{\mu^2 t_s^2}} \right)  ,
\ee
i.e. $\cV \sim e^{3a_s t_s/2}$ as expected. Note that only one solution is physical; the other corresponding to a negative volume.  In order to solve $\frac{\pd V}{\pd t_s} = 0$ for $t_s$, let us substitute (\ref{Vsol}) in
\be \label{Vdts}
\frac{\pd V}{\pd t_s} =  -\frac{ 2 a_s \lambda e^{- 2 a_s t_s}}{{\cal V}^{1/3}} + \frac{\mu (a_s t_s -1) e^{-a_s t_s}}{{\cal V}} \, .
\ee
It is convenient to use the approximation $a_s t_s \gg1$ as in \cite{BBCQ}, and thus the last equation becomes:
\be
-2 a_s \lambda e^{-2 a_s t_s} ({\cal V}^{2/3})^2 + \mu a_s t_s e^{-a_s t_s} {\cal V}^{2/3} = 0 \, .
\ee
Then, using the physical solution of (\ref{Vsol}), we obtain:
\be
1 + \sqrt{1 + \frac{20}{9} \frac{\lambda \nu}{\mu^2 t_s^2}} = \frac{1}{3} \, ,
\ee
which cannot be satisfied for any $t_s$.
This means that the minimization problem has no consistent solutions.

We just saw that $f(t) = f\left( \frac{t_s}{\cV^{1/3}} \right)$ has to be a nontrivial function in order to have a chance of stabilizing $\cV$ and $t_s$. It should be clear though, that such a non-trivial function would spoil the balance between the three contributions in (\ref{LVSPot}), due to the additional powers of $\cV$ that it introduces. To be more explicit in showing that there would be no large volume minima in this case, let us consider:
\bea
0 = t^\a\del_\a V = -{2\la a_s t_s e^{-2a_s t_s}\over \cV^{1/3}} + {\mu a_s t_s^2 e^{-a_s t_s} \over \cV} + {5\nu f(t)\over \cV^{5/3}} \, ,
\eea
where we recall that $t_\ell \sim \cV^{1/3}$ and we have used again $a_s t_s \gg1$. Therefore, at the minima the following relation is satisfied:
\bea \label{V}
\cV^{2/3} = {\mu t_s \over 4\la} e^{a_s t_s}\left(1+ \sqrt{1+40\left({\nu\la\over \mu^2}\right){f(t)\over a_s t_s^3}}\right).
\eea
Note, however, that this relation {\it does not} fix $\cV$ in terms of $t_s$, since $f(t)$ depends on $t_\ell \sim \cV^{1/3}$. Now, let us rewrite
\be
0 = \frac{\pd V}{\pd t_s} =  -\frac{ 2 a_s \lambda e^{- 2 a_s t_s}}{{\cal V}^{1/3}} + \frac{\mu (a_s t_s -1) e^{-a_s t_s}}{{\cal V}} - \frac{\nu f_{t_s}(t)}{{\cal V}^{5/3}} \, ,
\ee
as
\be \label{rewrite}
-2 a_s \lambda e^{-2 a_s t_s} ({\cal V}^{2/3})^2 + \mu a_s t_s e^{-a_s t_s} {\cal V}^{2/3} - \nu f_{t_s} = 0 \, ,
\ee
where again we have used $a_s t_s >\!\!> 1$. Plugging in \C{V} to eliminate the exponentials, we end up with the minimization condition
\be \label{min}
5 f(t) + t_s f_{t_s}(t ) =0 \,.
\ee
The important thing to note about \C{min} is that, given an $f$, this will fix $t_\ell\sim\cV^{1/3}$ in terms of $t_s$, but without any factors of $e^{a_s t_s}$.
So any values of $\cV$ and $t_s$ that satisfy this relation would be inconsistent with the large volume requirement that $\cV \sim e^{q a_s t_s}$ for some $q \in \mathbb{R}^{+}$.

\section{Three K\"{a}hler Moduli}
\setcounter{equation}{0}

In the previous section we saw that the case of two K\"{a}hler moduli is too restrictive and does not allow large volume minima. So let us now consider the next simplest possibility, namely three K\"{a}hler moduli. In this case, we will show that large volume minima can exist under certain conditions.

First, note the following main lessons we learned from the two moduli case: We needed a nontrivial function $f(t)$ and, in addition, we needed $f(t)$ to be independent of $\cV$. With only two moduli, these two conditions are inconsistent with each other, since $f$ is homogeneous of degree zero. However, with at least three moduli it is easy to satisfy both of them. So a necessary condition for the large volume minima we are after is that there are at least three K\"{a}hler moduli. Furthermore, let us assume that in the large volume limit:
\be
t_3 \sim \cV^{1/3} \gg t_1,t_2.
\ee
In other words, now we have two small moduli $t_{1,2}$ and generically ${\cal V} \sim t_{\ell}^3$, where we have denoted the large modulus $t_3 \equiv t_{\ell}$. Also, in accordance with the above discussion, we will assume that $f$ is of the form:
\be\label{f}
f(t) = f(t_1/t_2).
\ee
In this section, we will content ourselves with exploring the implications of \C{f} for moduli stabilization, leaving the important problem of demonstrating the existence of such solutions (and/or finding concrete examples) for future study. We should also note that this kind of situation, namely two small moduli and one large one, has been considered in detail, in the context of IIB orientifolds, in Appendix A.2 of \cite{CCQ}. We will begin by following their discussion. However, after a certain point there will be essential differences.

Since now we have two small moduli, the general expressions (\ref{Vnp1i}) and (\ref{Vnp2i}) acquire the following respective forms in the large volume limit:
\be
V_{np\,1} = e^K \sum_{\alpha=1}^{2} G^{\alpha \bar{\alpha}} a_{\alpha}^2 |A_{\alpha}|^2 e^{-2 a_{\alpha} t_{\alpha}} + \left( e^{K} G^{12} a_1 A_1 a_2 \bar{A}_2 e^{-a_1 t_1 - a_2 t_2} e^{i(a_2 b_2 - a_1 b_1)} + c.c.\right) \, ,
\ee
\be
V_{np\,2} = - e^K \sum_{\alpha=1}^{2} G^{\alpha \bar{\beta}} \left[ a_{\alpha} A_{\alpha} e^{-a_{\alpha} t_{\alpha}} e^{-i a_{\alpha} b_{\alpha}} \bar{W}_0 K_{\bar{\beta}} + c.c. \right] \, .
\ee
The first step in the minimization of the total scalar potential is again the elimination of the axions $b_{1,2}$.

\subsection{Axion minimization}

For convenience, let us denote the axion-dependent part of $V_{np\,1}$ by
\be
V_{np\,1}^{ax} = Y e^{i (a_2 b_2 -a_1 b_1)} + \bar{Y} e^{-i(a_2 b_2 -a_1 b_1)} \, ,
\ee
where
\be
Y \equiv e^K G^{12} a_1 a_2 A_1 \bar{A}_2 e^{-(a_1 t_1 + a_2 t_2)} \, .
\ee
Similarly, we introduce the notation
\be
V_{np\,2} = \sum_{\alpha=1}^{2} \left( X_{\alpha} e^{i a_{\alpha} b_{\alpha}} + \bar{X}_{\alpha} e^{-i a_{\alpha} b_{\alpha}} \right) \, ,
\ee
where
\be \label{Xdef}
X_{\alpha} \equiv - e^K G^{\bar{\alpha} \beta} K_{\beta} a_{\alpha} \bar{A}_{\alpha} W_0 e^{-a_{\alpha} t_{\alpha}} \, .
\ee
Now we can write the axion scalar potential $V_{ax} \equiv V_{np\,1}^{ax} + V_{np\,2}$ as:
\be
V_{ax} = 2 \sum_{\alpha=1}^2 |X_{\alpha}| \cos (\theta_W -\theta_{\alpha} + a_{\alpha} b_{\alpha}) + 2 |Y| \cos (\theta_1 - \theta_2 + a_2 b_2 - a_1 b_1) \, ,
\ee
where the $\theta$ angles are defined via
\be
W_0 = |W_0| e^{i \theta_W} \, , \qquad A_1 = |A_1| e^{i \theta_1} \, , \qquad A_2 = |A_2| e^{i \theta_2} \, .
\ee
The extremum conditions $\frac{\pd V_{ax}}{\pd b_1} = 0$ and $\frac{\pd V_{ax}}{\pd b_2} = 0$ can be solved by
\bea
&&\psi_1 \equiv \theta_W - \theta_1 + a_1 b_1 = n_1 \pi \, , \qquad n_1 \in \mathbb{Z} \\
&&\psi_2 \equiv \theta_W - \theta_2 + a_2 b_2 = n_2 \pi \, , \qquad n_2 \in \mathbb{Z} \, .
\eea
As a result, we also have that
\be
\psi_Y \equiv \theta_1 - \theta_2 + a_2 b_2 -a_1 b_1 = (n_2 - n_1) \pi \, .
\ee

In order to ensure a minimum of $V_{ax}$, we also need to look at its second derivatives. Let us compute the latter:
\bea
\frac{\pd^2 V_{ax}}{\pd b_1^2} &=& - 2 a_1^2 \left( |X_1| \cos \psi_1 + |Y| \cos \psi_Y \right) \, , \label{Minb1} \\
\frac{\pd^2 V_{ax}}{\pd b_2^2} &=& - 2 a_2^2 \left( |X_2| \cos \psi_2 + |Y| \cos \psi_Y \right) \, , \\
\frac{\pd^2 V_{ax}}{\pd b_1 \pd b_2} &=& 2 a_1 a_2 |Y| \cos \psi_Y \, .
\eea
The conditions for a minimum are
\be \label{Cond2d}
\frac{\pd^2 V_{ax}}{\pd b_1^2} \, \frac{\pd^2 V_{ax}}{\pd b_2^2} - \left( \frac{\pd^2 V_{ax}}{\pd b_1 \pd b_2} \right)^2 > 0 \qquad {\rm and} \qquad \frac{\pd^2 V_{ax}}{\pd b_1^2} > 0 \,\, .
\ee
These imply certain correlations between the signs of $\cos \psi_1$, $\cos \psi_2$ and $\cos \psi_Y$. In particular, one can immediately see from (\ref{Minb1}) that the second condition in (\ref{Cond2d}) can never be satisfied if both $\cos \psi_1$ and $\cos \psi_Y$ are positive. So we are left with the following three possibilities for a minimum:
\bea
\begin{array}{|c|r|r|r|}
\hline
 & (1) & (2) & (3) \\
\hline
\cos \psi_1 & -1 & 1 & -1 \\
\hline
\cos \psi_2 & 1 & -1 & -1 \\
\hline
\cos \psi_Y & -1 & -1 & 1 \\
\hline
\end{array} \label{Table}
\eea

In case (1) of (\ref{Table}), the first condition in (\ref{Cond2d}) acquires the form:
\be \label{Con1}
|X_1| |Y| > |X_2| \left( |X_1| + |Y| \right) \, ,
\ee
whereas the second condition is automatically satisfied.

In case (2) we have instead:
\be \label{Con2}
|X_2| |Y| > |X_1| \left( |X_2| + |Y| \right) \qquad {\rm and} \qquad |X_1| < |Y| \, .
\ee

Finally, in case (3) the conditions (\ref{Cond2d}) are:
\be \label{Con3}
|X_1| |X_2| > |Y| \left( |X_1| + |X_2| \right) \qquad {\rm and} \qquad |X_1| > |Y| \, .
\ee

In order to gain more insight into (\ref{Con1})-(\ref{Con3}), let us be more explicit about the quantities involved. From (\ref{Xdef}) and (\ref{G-inv}), together with ${\cal V}^{1 \alpha} {\cal V}_{\alpha} = \frac{1}{2} t_1$ and $t^{\alpha} {\cal V}_{\alpha} = 3 {\cal V}$, we find that
\be
|X_1| = a_1 |A_1| |W_0| \frac{t_1}{{\cal V}} e^{-a_1 t_1}
\ee
and similarly
\be
|X_2| = a_2 |A_2| |W_0| \frac{t_2}{{\cal V}} e^{-a_2 t_2} \, .
\ee
On the other hand, for $|Y|$ we have:
\be
|Y| = \left( - {\cal V}^{12} + \frac{t_1 t_2}{2 {\cal V}} \right) a_2 a_2 |A_1| |A_2| e^{-(a_1 t_1 + a_2 t_2)} \, .
\ee
So we need to know the large-${\cal V}$ behaviour of the inverse of ${\cal V}_{\alpha \beta} = \kappa_{\alpha \beta \gamma} t^{\gamma}$. Generically, it is given by ${\cal V}^{\alpha \beta} \sim {\cal V}^{-1/3}$ just as in the case of two moduli. Of course, other behaviours are possible when various sets of components of $\kappa_{\alpha \beta \gamma}$ vanish identically. In the following we will consider the generic case though, leaving the investigation of special cases for the future.

The minima we are looking for are of the type ${\cal V} \sim e^{p \,a_1 t_1} \sim e^{q \,a_2 t_2}$ with $p,q \in \R^{+}$. Therefore, at large volume:
\be \label{LVXY}
|X_1| \sim \frac{c_1}{{\cal V}^{1+\frac{1}{p}}} \,\, , \qquad |X_2| \sim \frac{c_2}{{\cal V}^{1+\frac{1}{q}}} \,\, , \qquad |Y| \sim \frac{c_Y}{{\cal V}^{\frac{1}{3}+\frac{1}{p}+\frac{1}{q}}} \,\, ,
\ee
where clearly the coefficients $c_{1,2,Y}$ are independent of ${\cal V}$, but do depend on the small moduli $t_1$ and $t_2$. Using (\ref{LVXY}), one can extract constraints on $p$ and $q$, such that a given condition in (\ref{Con1})-(\ref{Con3}) is satisfied regardless of the values of the coefficients $c_{1,2,Y}$. For example, the condition $|X_1| < |Y|$ is guaranteed to be true when $q > 3/2$. However, such a logic misses a set of minima for the following reason. If one only looks at the leading large-${\cal V}$ behaviour, then for $p=q$ one has that $|X_1| \sim |X_2|$. Denoting both $|X_{1,2}|$ by $|X|$, one then finds that the condition $|X| |Y| > |X| (|X|+|Y|)$, appearing in both (\ref{Con1}) and (\ref{Con2}), can never be satisfied as it reduces to $|X| < 0$, as was concluded in \cite{CCQ}. However, note that even when $|X_1| \sim |X_2|$, they may have different coefficients and so, for example, a condition of the form $|X_1| |Y| > |X_2| (|X_1|+|Y|)$ can reduce to $c_1 c_Y > c_2 ( c_1 + c_Y)$ for $p=q=\frac{3}{2}$. The latter condition may very well be possible to satisfy depending on the stabilization of the K\"{a}hler moduli and on the choice of values for the various parameters. So for the moment we will leave further analysis of the axion minimum conditions until after we have considered the stabilization of ${\cal V}$, $t_1$ and $t_2$. Once we have established the conditions arising from the K\"{a}hler moduli stabilization, we will come back to the axions and see whether it is possible to satisfy all minimization conditions at the same time.

\subsection{K\"{a}hler moduli potential}

Let us now consider the scalar potential for the K\"{a}hler moduli that results from stabilizing the axions. In the following, for definiteness we take  case (3) of (\ref{Table}); clearly, the other two cases amount to just sign changes in front of some of the terms and so we can easily take them into account at the end. Therefore, we have the scalar potential
\be
V = V_{np\,1}^{real} + 2 |Y| - 2 |X_1| - 2 |X_2| + V_{\alpha'} \,\, ,
\ee
where
\be
V_{np\,1}^{real} = e^K \sum_{\alpha=1}^2 G^{\alpha \bar{\alpha}} a_{\alpha}^2 |A_{\alpha}|^2 e^{-2 a_{\alpha} t_{\alpha}} \qquad {\rm and} \qquad V_{\alpha'} = - \frac{\nu f (t_1/t_2)}{{\cal V}^{5/3}} \,\, .
\ee
For convenience, let us introduce notation similar to the two modulus case. Namely, we denote:
\be
V_{np\,1}^{real} = \sum_{\alpha=1}^2 \left( - {\cal V}^{\alpha \alpha} + \frac{t_{\alpha}^2}{2 {\cal V}} \right) a_{\alpha}^2 |A_{\alpha}|^2 e^{-2 a_{\alpha} t_{\alpha}} \equiv \sum_{\alpha=1}^2 \frac{\lambda_{\alpha}}{{\cal V}^{1/3}} e^{-2 a_{\alpha} t_{\alpha}} + ... \, ,
\ee
where we have used that generically at large volume ${\cal V}^{\alpha \alpha} \sim {\cal V}^{-1/3}$ and, further, the constants $\lambda_{1,2}$ are positive just like $\lambda$ in (\ref{ldef}). Similarly, we introduce the notation
\be
2 |X_{\alpha}| = \mu_{\alpha} \frac{t_{\alpha}}{{\cal V}} e^{-a_{\alpha} t_{\alpha}} \qquad {\rm with} \qquad \mu_{\alpha} = 2 a_{\alpha} |A_{\alpha}| |W_0|
\ee
and
\be
2|Y| = \frac{\sigma}{{\cal V}^{1/3}} e^{-(a_1 t_1 + a_2 t_2)} \, ,
\ee
where again $\mu_{1,2}$ and $\sigma$ are positive constants.

Hence, the scalar potential acquires the form:
\be \label{LVSPot2}
V = \lambda_1 \frac{e^{-2 a_1 t_1}}{{\cal V}^{1/3}} + \lambda_2 \frac{e^{-2 a_2 t_2}}{{\cal V}^{1/3}} + \sigma \frac{e^{-a_1 t_1} e^{-a_2 t_2}}{{\cal V}^{1/3}} - \mu_1 \frac{t_1}{{\cal V}} e^{-a_1 t_1} - \mu_2 \frac{t_2}{{\cal V}} e^{-a_2 t_2} - \nu \frac{f(\frac{t_1}{t_2})}{{\cal V}^{5/3}} \, .
\ee
Clearly, this potential has structure very similar to the one of (\ref{LVSPot}). And as in that previous case, now too it is obvious that all terms in (\ref{LVSPot2}) will be of the same order of magnitude only for ${\cal V} \sim e^{\frac{3}{2} a_1 t_1} \sim e^{\frac{3}{2} a_2 t_2}$. So we will take $p$ and $q$ in (\ref{LVXY}) to be $p=q=\frac{3}{2}$. Of course, it could be possible that some of the terms in (\ref{LVSPot2}) are subleading and can be neglected, while the rest balance each other so that there is a minimum for ${\cal V}$ at a large value. However, in such a case it seems unlikely that both small moduli $t_{1,2}$ can be stabilized by the leading potential. So we will not consider this possibility further at this time.

Let us now turn to the minimization conditions. The condition $\pd V / \pd {\cal V} = 0$ can be solved by
\bea \label{Vmin3m}
{\cal V}^{2/3} &=& \frac{3}{2} \frac{\mu_1 t_1 e^{-a_1 t_1} + \mu_2 t_2 e^{-a_2 t_2}}{\lambda_1 e^{-2 a_1 t_1} + \lambda_2 e^{-2 a_2 t_2} + \sigma e^{-(a_1 t_1 + a_2 t_2)}} \times \nn\\ \nn\\
&\times& \left( 1 \pm \sqrt{1+ \frac{20}{9} \nu f \frac{\lambda_1 e^{-2 a_1 t_1}+\lambda_2 e^{-2 a_2 t_2}+ \sigma e^{-(a_1 t_1 + a_2 t_2)}}{(\mu_1 t_1 e^{-a_1 t_1}+\mu_2 t_2 e^{-a_2 t_2})^2}} \right) ,
\eea
which for $e^{a_1 t_1} \sim e^{a_2 t_2}$ indeed gives ${\cal V}^{2/3}\sim e^{a_1 t_1}$. Ideally, we would want now to use
\be \label{Mint1}
\frac{\pd V}{\pd t_1} = -\frac{2 a_1 \lambda_1 e^{-2 a_1 t_1}}{{\cal V}^{1/3}} + \frac{\mu_1 a_1 t_1 e^{-a_1 t_1}}{{\cal V}} - \frac{\sigma a_1 e^{-(a_1 t_1 + a_2 t_2)}}{{\cal V}^{1/3}} - \frac{\nu f_{t_1}}{{\cal V}^{5/3}} = 0
\ee
and
\be \label{Mint2}
\frac{\pd V}{\pd t_2} =  -\frac{2 a_2 \lambda_2 e^{-2 a_2 t_2}}{{\cal V}^{1/3}} + \frac{\mu_2 a_2 t_2 e^{-a_2 t_2}}{{\cal V}} - \frac{\sigma a_2 e^{-(a_1 t_1 + a_2 t_2)}}{{\cal V}^{1/3}} - \frac{\nu f_{t_2}}{{\cal V}^{5/3}} = 0
\ee
in order to solve for $t_{1,2}$ (here again we have used $a_1 t_1 >\!\!> 1$ and $a_2 t_2 >\!\!> 1$). However, due to the presence of both exponentials and powers of $t_{1,2}$ and, furthermore, of mixed terms, it is not possible to write down an explicit analytic solution. One can still convince oneself numerically that for appropriate choices of parameters the system of equations $\pd V / \pd {\cal V} = 0$ , $\pd V / \pd t_1 = 0$, $\pd V / \pd t_2 = 0$ has consistent solutions, due to the fact that $f_{t_1}$ and $f_{t_2}$ have opposite signs since $f=f(\frac{t_1}{t_2})$. However, in order to gain understanding as to what goes differently compared to the two modulus case, in which the equation $\pd V / \pd t_s = 0$ was not possible to solve for $t_s \in \mathbb{R}^{+}$, let us now make some simplifications that will allow us to analyze (\ref{Mint1})-(\ref{Mint2}) analytically.

First of all, note that since we are considering the case $e^{a_1 t_1} \sim e^{a_2 t_2}$, so that we can stabilize both $t_1$ and $t_2$, then we can write $e^{a_1 t_1} = c e^{a_2 t_2}$ at the minimum. Here the constant $c$ can depend on the stabilized values of $t_{1,2}$ and the various parameters. The precise dependence is not of paramount importance though; we are just trying to make an analytical estimate of various relevant terms. Now, using this relation between the two exponentials, we can rewrite (\ref{Vmin3m}) as:
\bea \label{Vminc}
{\cal V}^{2/3} &=& \frac{3}{2} e^{a_1 t_1} \frac{\mu_1 t_1 + \mu_2 c t_2}{\lambda_1 + \lambda_2 c^2 + \sigma c} \left( 1 \pm \sqrt{1+ \frac{20 \nu f}{9} \frac{(\lambda_1 + \lambda_2 c^2 + \sigma c)}{(\mu_1 t_1 + \mu_2 c t_2)^2}} \right) \nn\\
&=& \frac{3}{2} e^{a_2 t_2} c \frac{\mu_1 t_1 + \mu_2 c t_2}{\lambda_1 + \lambda_2 c^2 + \sigma c} \left( 1 \pm \sqrt{1+ \frac{20 \nu f}{9} \frac{(\lambda_1 + \lambda_2 c^2 + \sigma c)}{(\mu_1 t_1 + \mu_2 c t_2)^2}} \right) .
\eea
Using the first line of (\ref{Vminc}), we can rewrite (\ref{Mint1}) as:
\be \label{Mint1r}
\frac{3}{2} \frac{\mu_1 t_1 + \mu_2 c t_2}{\lambda_1 + \lambda_2 c^2 + \sigma c} \left( 1 \pm \sqrt{1+ \frac{20 \nu f}{9} \frac{(\lambda_1 + \lambda_2 c^2 + \sigma c)}{(\mu_1 t_1 + \mu_2 c t_2)^2}} \right) = \frac{\mu_1 a_1 t_1 \pm \sqrt{P}}{2 a_1 (2 \lambda_1 + \sigma c)} \,\, ,
\ee
where
\be
P = \mu_1^2 a_1^2 t_1^2 - 4 a_1 (2 \lambda_1 + \sigma c) \nu f_{t_1} \,\, .
\ee
Similarly, using the second line of (\ref{Vminc}), we can rewrite (\ref{Mint2}) as:
\be \label{Mint2r}
\frac{3}{2} c \frac{\mu_1 t_1 + \mu_2 c t_2}{\lambda_1 + \lambda_2 c^2 + \sigma c} \left( 1 \pm \sqrt{1+ \frac{20 \nu f}{9} \frac{(\lambda_1 + \lambda_2 c^2 + \sigma c)}{(\mu_1 t_1 + \mu_2 c t_2)^2}} \right) = \frac{\mu_2 a_2 t_2 \pm \sqrt{Q}}{2 a_2 (2 \lambda_2 + \frac{\sigma}{c})} \,\, ,
\ee
where
\be
Q = \mu_2^2 a_2^2 t_2^2 - 4 a_2 \left( 2 \lambda_2 + \frac{\sigma}{c} \right) \nu f_{t_2} \,\, .
\ee
Comparing (\ref{Mint1r}) and (\ref{Mint2r}) implies:
\be \label{CompEq}
\frac{\mu_1 a_1 t_1 \pm \sqrt{P}}{2 a_1 (2 \lambda_1 + \sigma c)} = \frac{\mu_2 a_2 t_2 \pm \sqrt{Q}}{2 a_2 (2 \lambda_2 c + \sigma)} \,\, .
\ee

Clearly, solving the pair of equations (\ref{Mint1r}) and (\ref{Mint2r}) is equivalent to solving (\ref{CompEq}) together with, say, (\ref{Mint1r}). However, this last system is still not very illuminating. So let us make a further simplification, that will enable us to handle things analytically. Namely, we take $c=1$, which implies that at the minimum $a_1 t_1 = a_2 t_2$. Note that, in fact, this is not a strong restriction at all since any small difference between $a_1 t_1$ and $a_2 t_2$ will be greatly magnified by the exponentiation. So if $e^{a_1 t_1}$ and $e^{a_2 t_2}$ are to be of the same order of magnitude in the region of moduli space that we are studying, then the numerical difference between $a_1 t_1$ and $a_2 t_2$ cannot be very significant. Now, we can use the relation $a_1 t_1 = a_2 t_2$ and one of (\ref{CompEq}), (\ref{Mint1r}) to solve for $t_{1,2}$ at the minimum and view the remaining relation as a constraint among the parameters.\footnote{This may not seem very satisfactory. However, remember that we are not trying here to solve things in full generality, which is only possible numerically. We are just looking for an analytically manageable special case, that gives a consistent solution, in order to show in principle that it is possible to have large volume minima. Such a special case may mean a particular choice for the function $f(t_1/t_2)$, but also some constraints/relations among the parameters. All we need to do, for our purposes, is to show at the end that it is possible to satisfy those constraints. In contrast, recall that for the two moduli case there was {\it no} choice of parameters that would give a real positive result for $t_s$.}

To make further progress analytically, we need to take a concrete function $f(t_1/t_2)$. The simplest possibility, namely a constant, can be shown to give no consistent solution just like for the case of two K\"{a}hler moduli. So we need a nontrivial function. A convenient choice turns out to be the following:
\be \label{fdef}
f (t_1 / t_2) = \frac{t_1^3}{t_2^3} \, .
\ee
Hence
\be
f_{t_1} = 3 \frac{t_1^2}{t_2^3} \qquad {\rm and} \qquad f_{t_2} = - 3 \frac{t_1^3}{t_2^4} \, .
\ee
Substituting these derivatives in (\ref{CompEq}), and then using $a_1 t_1 = a_2 t_2$, we find
\be
\frac{\mu_1 a_1 + \sqrt{\mu_1^2 a_1^2 - 12 \nu a_1 (2 \lambda_1 + \sigma)\frac{1}{t_2^3}}}{2 a_1 (2 \lambda_1 + \sigma)} = \frac{\mu_2 a_1 + \sqrt{\mu_2 a_1^2+ 12 \nu a_2 (2 \lambda_2 + \sigma) \frac{a_2}{a_1} \frac{1}{t_2^3}}}{2 a_2 (2 \lambda_2 + \sigma)}
\ee
which can easily be solved:
\be \label{t2Sol}
\frac{1}{t_2^3} = \frac{a_1 (a_1 \mu_2 + a_2 \mu_1)}{12 \,\nu \,a_2^2 (\lambda_1+\lambda_2 + \sigma)^2} \left[ a_2 \mu_1 (2 \lambda_2 + \sigma) - a_1 \mu_2 (2 \lambda_1 + \sigma ) \right]
\ee
Clearly, for a consistent solution, we need that the following inequality be satisfied:
\be \label{Ineq1}
a_2 \mu_1 (2 \lambda_2 + \sigma) > a_1 \mu_2 (2 \lambda_1 + \sigma ) \, .
\ee
In addition, we are still left with one more equation, namely (\ref{Mint1r}). Let us see what expression for $t_{1,2}$ follows from it. For convenience, we introduce the notation:
\be \label{Nu}
\nu = K t_2^3 \, ,
\ee
where the constant $K$ is defined via (\ref{t2Sol}). Now, substituting (\ref{Nu}) and $t_1 = \frac{a_2}{a_1} t_2$ in (\ref{Mint1r}) and using (\ref{fdef}), we obtain:
\be
\frac{3 (\mu_1 a_2 + \mu_2 a_1)}{a_2 (\lambda_1 + \lambda_2 + \sigma)} \!\left( 1 + \sqrt{1 + \frac{20}{9} \frac{a_2^2 K (\lambda_1 + \lambda_2 + \sigma)}{(\mu_1 a_2 + \mu_2 a_1)^2} t_1}  \right) \!= \frac{\mu_1 a_1 + \sqrt{\mu_1^2 a_1^2 -12 K a_1 (2 \lambda_1 + \sigma)}}{2 a_1 (2 \lambda_1 + \sigma)} \, ,
\ee
which is also easy to solve:
\be \label{t1Sol}
t_1 = \frac{\lambda_1 + \lambda_2 + \sigma}{ 5 K } \,C \left( C - 3 \frac{\mu_1 a_2 + \mu_2 a_1}{ a_2 (\lambda_1 + \lambda_2 + \sigma)}\right) \, ,
\ee
where
\be
C = \frac{\mu_1 a_1 + \sqrt{\mu_1^2 a_1^2 -12 K a_1 (2 \lambda_1 + \sigma)}}{2 a_1 (2 \lambda_1 + \sigma)} \,\, .
\ee

To recapitulate, we have solved the minimization conditions for the small moduli $t_{1,2}$. In order for the latter to be real positive numbers, the parameter values have to be chosen such that several inequalities are satisfied. One is (\ref{Ineq1}) and another two follow from (\ref{t1Sol}). Namely:
\be
\mu_1^2 a_1^2 > 12 K a_1 (2 \lambda_1 + \sigma) \qquad {\rm and} \qquad C > 3 \frac{\mu_1 a_2 + \mu_2 a_1}{ a_2 (\lambda_1 + \lambda_2 + \sigma)} \, .
\ee
Also, the relation
\be
\nu = K \frac{a_1^3}{a_2^3} t_1^3 \, ,
\ee
with $t_1$ substituted from (\ref{t1Sol}), has to be viewed as a constraint between the parameters.

Finally, recall that we also have to satisfy the axion minimization conditions. Since we have been considering case (3) in (\ref{Table}) and we have $p=q=3/2$ in (\ref{LVXY}), the relevant axion conditions (\ref{Con3}) acquire the form:
\be \label{axCond}
\mu_1 t_1 > \sigma \qquad {\rm and} \qquad \mu_1 t_1 \mu_2 t_2 > \sigma (\mu_1 t_1 + \mu_2 t_2) \, .
\ee
Note that, upon using $a_1 t_1 = a_2 t_2$, the second condition becomes:
\be
\mu_1 t_1 > \sigma \left( 1+ \frac{\mu_1}{\mu_2} \frac{a_2}{a_1} \right) \, .
\ee
So whenever it is satisfied, then the first condition in (\ref{axCond}) follows automatically.

It may seem that there are quite a few constraints on the parameters. However, one can easily find various examples that satisfy all conditions above. For instance, the set of parameter values $\{$ $\lambda_1 = 1$, $\mu_1 = 6$, $a_1 = 8$, $\lambda_2 = 10$, $\mu_2 = 1$, $a_2=9$, $\sigma = 2$ $\}$ gives $t_1 = 61.02$ and $t_2 = 54.24$, i.e. the small moduli are stabilized not only at real positive values but also at values that are large enough for the supergravity approximation to be reliable.

\subsection*{Acknowledgements}

We would like to thank M. Cicoli, J. Conlon, A. Parnachev, N. Saulina and S. Sethi for useful conversations. L.A. is supported by DOE grant FG02-84-ER40153. C.Q. is
supported in part by NSF Grant No. PHY-0758029 and by an NSERC PGS-D Scholarship.

\appendix

\section{On Special Hermitian Manifolds} \label{SpHerm}
\setcounter{equation}{0}

Special Hermitian manifolds are $SU(3)$ structure manifolds, for which $dJ \wedge \Omega = 0$ identically. To explain that, let us review a few facts about manifolds with $SU(3)$ structure. (For more details see \cite{CCDLMZ}.)
The latter are characterized by five torsion classes $W_i$ with $i=1,...,5$, which determine the deviation from $SU(3)$ holonomy. More precisely, $W_i$ are defined via:
\bea \label{tordef}
dJ &=& \frac{3}{4} \,i \left( W_1 \bar{\Omega} - \bar{W}_1 \Omega \right) + W_3 + J\wedge W_4  \nn \\
d\Omega &=& W_1 J\wedge J + J\wedge W_2 + \Omega \wedge W_5  ,
\eea
where
\be \label{compat}
J \wedge W_3 =  0 \, , \qquad \Omega \wedge W_3 = 0 \qquad {\rm and} \qquad J\wedge J \wedge W_2 = 0  .
\ee
The last three conditions are necessary to guarantee the independence of the various terms in (\ref{tordef}). For example, the condition $J\wedge J\wedge W_2 = 0$ implies that the $J\wedge W_2$ term does not contain a piece proportional to $J\wedge J$. Let us also note that the appearance of the same torsion class, $W_1$, in both the $dJ$ and $d\Omega$ equations is due to (\ref{compat}) and the $SU(3)$ structure compatibility conditions:
\be \label{SU3comp}
J\wedge \Omega = 0 \qquad {\rm and} \qquad J \wedge J \wedge J = \frac{3}{4} \,i \,\Omega \wedge \bar{\Omega}  .
\ee
The various kinds of $SU(3)$ structure manifolds are classified by the vanishing of certain subsets of torsion classes.\footnote{For example, when all $W_i$ vanish one has a CY 3-fold.}

Now, the heterotic susy conditions imply that the internal manifold has to be complex and therefore for us $W_{1,2} = 0$. Another consequence of the supersymmetry conditions for the heterotic string is that $2W_4 = - W_5$ \cite{CCDLMZ}. Hence, vanishing $W_4$ implies $W_5 = 0$ and vice versa. As a result, we see from (\ref{tordef}) that if $dJ = 0$ (i.e., if $W_3 = 0$ and $W_4 = 0$), then also $d\Omega = 0$ and we are back to considering CY 3-folds. So, in order to have an $SU(3)$ structure manifold that takes into account the backreaction of the $H$-flux, we need to have $dJ \neq 0$. This together with the first compatibility condition in (\ref{SU3comp}), implies that we can ensure $dJ \wedge \Omega = 0$ by taking $d\Omega =0$ (i.e., $W_5 = 0$). Therefore, we are led to consider manifolds, whose only nonvanishing torsion class is $W_3$. (On shell $W_3$ is given by the $H$-flux.) Such manifolds are called {\it special Hermitian}.

Finally, let us also note that compactifications on special Hermitian manifolds have constant dilaton $\phi$, since $d \phi$ is proportional to the vanishing torsion class $W_4$ (or, equivalently, $W_5$) \cite{CCDLMZ}.

\section{The No-Scale Breaking Potential}\label{NoScaleBreaking}
\setcounter{equation}{0}

Here we give the calculation of the leading $\ap$ correction to the scalar potential. First note that, using (\ref{RelVt}), one can derive the following useful relations:
\bea
&& \cV_\a t^\a = 3\cV \\
&& \cV_{\a\b} t^\b = 2 \cV_\a \\
&& \cV^{\a\b}\cV_\b = \hlf \cV^{\a\b}\cV_{\b\g}t^\g = \hlf t^\a
\eea

Now, let us start with the following (corrected) K\"ahler potential:
\be
K = -\log(\cV+\D \cV) = -\log(\cV) -{\D \cV\over \cV} +O(\D \cV^2) \, ,
\ee
where for the time being we leave $\D \cV$ unspecified for convenience; at the end we will take
$\D \cV={\ap\over2}\int J\wedge h\wedge h$. We will always work to leading
order in $\D \cV$ only, dropping all higher order terms. It is straightforward to compute:
\be
K_\a = -{\cV_\a\over \cV}\Big(1-{\D \cV\over \cV}\Big) - {\D \cV_\a\over \cV}
\ee
and
\be
K_{\a\b} = K_{\a\b}\0\Big(1-{\D \cV\over \cV}\Big) - {N_{\a\b}\over \cV} \, ,
\ee
where
\be
K_{\a\b}\0 = -{\cV_{\a\b}\over \cV} + {\cV_\a \cV_\b \over \cV^2}
\ee
is the classical metric, and
\be
N_{\a\b} = \D \cV_{\a\b} - {\D \cV_\a \cV_\b + \D \cV_\b \cV_\a \over \cV} + {\D \cV \cV_\a \cV_\b\over \cV^2} \, .
\ee
It is easy to find the inverse metric, at least to $O(\D \cV)$:
\be
K^{\a\b} = K\zero{\a\b}\Big(1-{\D \cV\over \cV}\Big)^{-1} + {N^{\a\b}\over \cV} +O(\D \cV^2) \, ,
\ee
where
\be
K\zero{\a\b} = -\cV \cV^{\a\b} +\hlf t^\a t^\b
\ee
is the classical inverse metric used to raise and lower indices:
\be
N^{\a\b} = K\zero{\a\g}K\zero{\b\dd}N_{\g\dd} \, .
\ee

Now let us examine the no-scale breaking term:
\be
K^{\a\b}K_\a K_\b = {K\zero{\a\b}\cV_\a \cV_\b \over \cV^2}\Big(1-{\D \cV\over \cV}\Big)
+ 2{K\zero{\a\b}\cV_\a \D \cV_\b \over \cV^2} + {N^{\a\b}\cV_\a \cV_\b \over \cV^3} +O(\D \cV^2) \, .
\ee
Using the relation $K\zero{\a\b}\cV_\a = \cV t^\a$, this simplifies to:
\bea
K^{\a\b}K_\a K_\b -3 &=& 3\Big(1-{\D \cV\over \cV}\Big) + {2\over \cV}t^\a \D \cV_\a
+ {t^\a t^\b N_{\a\b}\over \cV} -3\\
&=&-3{\D \cV\over \cV} + {2\over \cV}t^\a \D \cV_\a
+ {1\over \cV}\big( t^\a t^\b \D \cV_{\a\b} -6t^\a \D \cV_\a + 9\D \cV \big)  \\
&=& {1\over \cV} \big(t^\a t^\b \D \cV_{\a\b} - 4t^\a \D \cV_\a + 6\D \cV \big) \, . \label{noscale}
\eea
At this point, let us take:
\bea
\D \cV = {\ap^2\over2} t^\a f_\a(t)
\eea
with $f_\a=\int\o_\a\wedge h\wedge h$ being homogeneous degree zero functions, as in the main text.
Then, we have:
\bea
t^\a \D \cV_\a = {\ap^2\over2}t^\a\left(f_\a + t^\b f_{\b,\a} \right) = \D \cV \\
t^\a t^\b \D \cV_{\a\b} = {\ap^2\over2}t^\a t^\b \left(f_{\a,\b}+f_{\b,\a} +t^\g f_{\g,\a\b}\right)=0 \, ,
\eea
where $f_{\a,\b}=\del_\b f_\a$ and we have used that $t^\b f_{\a,\b}=0$ since $f_\a$ are degree zero. Hence, (\ref{noscale}) acquires the form:
\be
K^{\a\b}K_\a K_\b -3  = 2{\D \cV\over \cV} = {\ap^2\over \cV}t^\a f_\a(\hat{t})
= {\ap^2\over \cV}\int J \wedge h \wedge h\leq0 \, ,
\ee
where we have used that $\int J \wedge h \wedge h=-(h,h)\leq0$. This then gives
the following contribution to the scalar potential:
\be
V_{\ap} = -\ap^2 {|W_0|^2\over \cV^{2}}(h,h) \, .
\ee



\end{document}